\documentclass[aps,prl,preprint,showpacs,amsmath,floatfix,superscriptaddress]{revtex4}

\usepackage{graphicx}

\begin{document}

\title{Tunable Band Gap and Anisotropic Optical Response in Few-layer Black Phosphorus}

\author{Vy Tran}

\affiliation{Department of Physics, Washington University in St.
Louis, St. Louis, MO 63136, USA}

\author{Ryan Soklaski}

\affiliation{Department of Physics, Washington University in St.
Louis, St. Louis, MO 63136, USA}

\author{Yufeng Liang}

\affiliation{Department of Physics, Washington University in St.
Louis, St. Louis, MO 63136, USA}

\author {Li Yang}

\affiliation{Department of Physics, Washington University in St.
Louis, St. Louis, MO 63136, USA}

\date{\today}

\begin{abstract}
We report the quasiparticle band gap, excitons, and highly
anisotropic optical responses of few-layer black phosphorous
(phosphorene). It is shown that these new materials exhibit unique
many-electron effects; the electronic structures are dispersive
essentially along one dimension, leading to particularly enhanced
self-energy corrections and excitonic effects. Additionally,
within a wide energy range, including infrared light and part of
visible light, few-layer black phosphorous absorbs light polarized
along the structure's armchair direction and is transparent to
light polarized along the zigzag direction, making them viable
linear polarizers for applications. Finally, the number of
phosphorene layers included in the stack controls the material's
band gap, optical absorption spectrum, and anisotropic
polarization energy-window across a wide range.
\end{abstract}

\maketitle

It is difficult to overstate the interest in graphene and
graphene-inspired two-dimensional (2D) crystals \cite{2004MonoG,
2007Graphene, review-1}. Recently, an attractive finite-gapped 2D
semiconductor, few-layer black phosphorus (phosphorene), has been
successfully fabricated \cite{2014zhang, 2014ye, 2014jia, new-1}.
Despite phosphorene's promising direct band gap, one cannot
realistically envision phosphorene devices until its fundamental
excited-state properties, such as its quasiparticle (QP) band gap
and optical spectrum, are obtained. No such experimental
measurements have been made, thus corresponding first-principles
predictions are indispensable. Furthermore, these excited-state
quantities are known to be dictated by many-electron effects,
\emph{e.g.}, electron-electron (\emph{e-e}) and electron-hole
(\emph{e-h}) interactions. Therefore, it is essential to turn to
\emph{ab-initio} calculations that incorporate many-body
self-energy corrections and excitonic effects to advance the
forefront of phosphorene research and its applications.

In addition to its direct band gap, few-layer black phosphorus
exhibits unique excitonic effects that have not been observed in
other 2D structures. For example, recent experiments have observed
a strongly anisotropic conducting behavior \cite{2014ye}. It is of
particular interest to investigate the effect that this anisotropy
has on exciton formation and thus on the excited-state properties
of the material. Van der Waals (vdW) interactions allow layers of
phosphorene to be stacked; the band gap of the resulting material
depends on the number of stacked layers ($N$) \cite{2014ye,
2014wei}. This has been observed in other 2D semiconductors
\cite{2010wang, 2012gross, 2013jcp}. Because the band gap is a
crucial factor in determining electronic screening and
corresponding many-electron interactions in the material, the
optical spectra and excitonic effects of few-layer phosphorene can
also be controlled by the number of stacking layers. Accordingly,
studying few-layer phosphorene provides a chance to observe how
the electronic structure and excitonic properties of a 2D material
transition to that of a 3D material.

In this Letter, we perform first-principles GW-Bethe-Salpeter
Equation (BSE) simulations to study the QP band gap and optical
spectra of few-layer and bulk black phosphorous, resulting in
several important findings. First, we observe significant
many-electron effects. For monolayer phosphorene, the self-energy
correction enlarges the band gap from 0.8 eV to 2 eV and the
lowest-energy optical absorption peak is reduced to 1.2 eV because
of a huge exciton binding energy (around 800 meV). These strong
many-electron effects result from unique quasi-one dimensional
(1D) band dispersions of phosphorene. Second, we observe highly
anisotropic optical responses. Few-layer black phosphorous
strongly absorbs light polarized along its lattice's armchair
direction, but it transparent to light polarized along the zigzag
direction. It is chiefly absorbant across the infrared-light range
and part of the visible-light range, making it an ideal candidate
for being used as an optical linear polarizer with a wide energy
window. Finally, the band gap, exciton binding energies, optical
absorption spectrum, and linear polarization energy window of
phosphorene can all be broadly tuned by changing the number of
stacked layers. This serves as a convenient and efficient method
for engineering the material's excited-state properties.

\begin{figure}
\includegraphics*[scale=0.40]{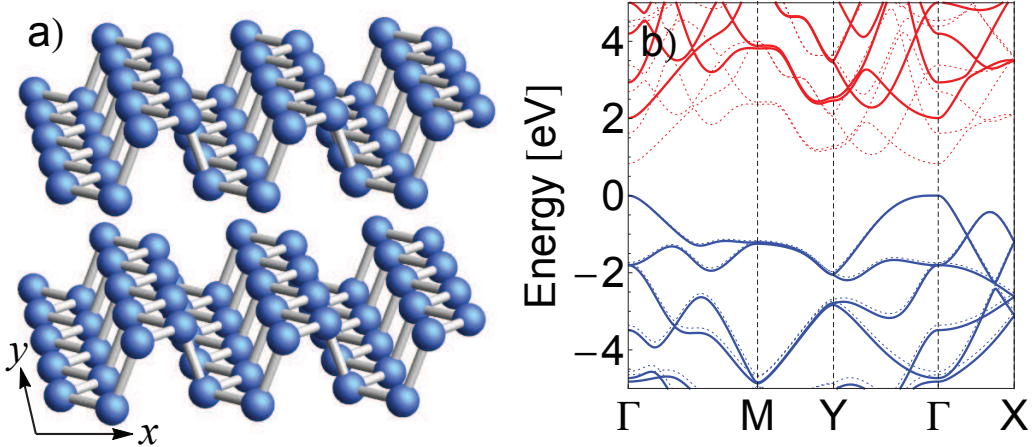}
\caption{(Color online) (a) The ball-stick model of few-layer
phosphorous. The x axis is perpendicular to and the y direction
parallel with the ridge direction, as indicated. (b) The
DFT-calculated (dash lines) and GW-calculated (solid lines) band
structures of monolayer phosphorous. The top of valence band is
set to be zero.} \label{fig:1}
\end{figure}

The atomistic ball-stick models of few-layer black phosphorus are
presented in Fig. \ref{fig:1} (a) \cite{struc-1, struc-2}. They
are fully relaxed according to the force and stress calculated by
density functional theory (DFT) within the PBE functional
\cite{pbe}. The ground-state wave functions and eigenvalues are
calculated by DFT/PBE with a k-point grid of 14x10x1 for few-layer
structures and 14x10x4 for bulk structures. All calculations use a
plane-wave basis with a 25 Ry energy cutoff with a norm-conserving
pseudopotential \cite{pseudo}. The QP energy is calculated by the
single-shot G$_0$W$_0$ approximation with the general plasmon pole
model \cite{1986Hybertsen}. The involved unoccupied band number is
about 10 times of that of valence bands to achieve the converged
dielectric function. The excitonic effects are included by solving
the BSE with a finer k-point grid of 56x40x1 (35x25x10 for bulk
structures) \cite{2000Rohlfing}. A slab Coulomb truncation is
crucial to mimic suspended structures \cite{trunc-1, trunc-2}.
Because of the depolarization effect \cite{dp-1, 2007yang}, only
the incident light polarized parallel with the plane structure is
considered in studying optical spectra.

The DFT and QP band structures of monolayer phosphorene are
presented in Fig. \ref{fig:1} (b). The band gap is located at the
$\Gamma$ point and the self-energy correction enlarges the band
gap from the DFT value of 0.8 eV to 2.0 eV, which is ideal for
broad electronic applications \cite{2011kis, 2013das}. The
calculated 150$\%$ enhancement is substantially larger than in
other 2D semiconductors \cite{BN-1, BN-2, MoS2-1, MoS2-2}. This is
due to the highly anisotropic band structure of the lowest
conduction band and highest valence band, as shown in Fig.
\ref{fig:1} (b). In particular, the band dispersion is very flat
along the $\Gamma-Y$ (zigzag) direction, confining particles to an
effective 1D environment along the armchair direction. This
effective lower dimension contributes to a larger self-energy
correction. The Coulomb truncation is crucial in our calculations
of suspended samples. Otherwise, the QP band gap will be
substantially smaller, \emph{e.g.}, the monolayer QP band gap will
be 1.7 eV. The worse case is that this value fluctuates by
different vacuum spacings in simulations.

\begin{figure}
\includegraphics*[scale=0.40]{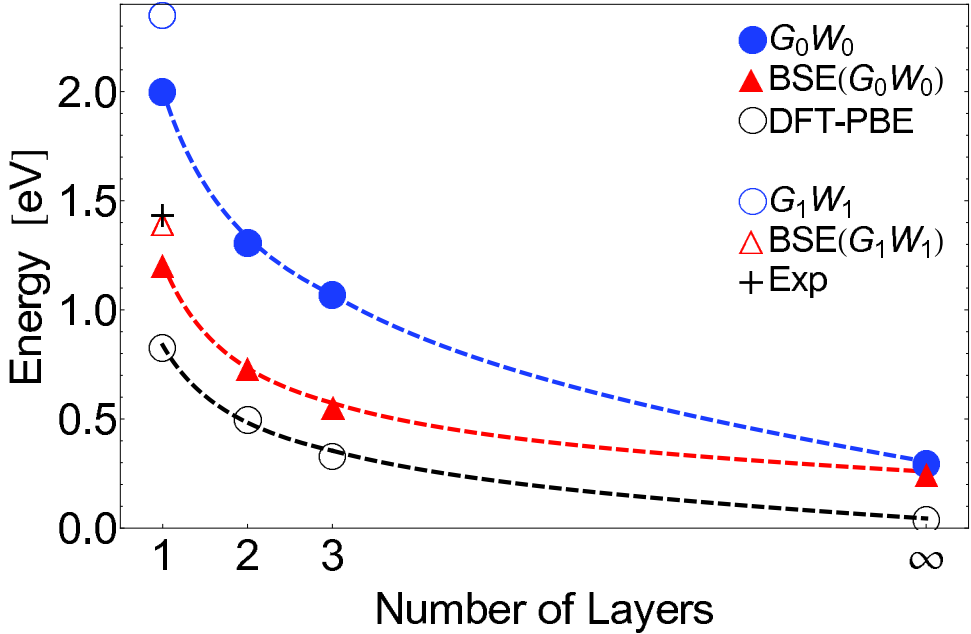}
\caption{(Color online) The evolution of band gap calculated by
different methods and optical absorption peak according to the
stacking layer number of few-layer phosphorene. The power-law
fitting curves are presented by dashed lines. The experimental
optical peak position is read from Ref. \cite{2014ye}.}
\label{fig:2}
\end{figure}

For multilayer phosphorene, the QP band gap and self-energy
corrections vary dramatically according to the stacking layer
number, although their band-structure topologies are roughly
similar to that of monolayer. As shown in Fig. \ref{fig:2}, the QP
band gap changes from 2 eV (monolayer) to 0.3 eV (bulk). These
bounds provide a wide range of tunability for the band gap and
corresponding electronic properties. As an evidence of the
reliability of our simulation, our calculated QP band gap of bulk
phosphorous is around 300 meV, which is in excellent agreement
with experimental results (0.33 eV) \cite{transport-1,
transport-2}.

A power law fit of the form ($A/N^{\alpha}+B$), where $N$ is the
number of layers, is applied in Fig. \ref{fig:2}, the results of
which are reported in Table I. Surprisingly, the GW-calculated
band gaps follows the $1/N^{0.7}$ power law, which decays
significantly slower than the usual quantum confinement result
($1/N^2$). This weaker quantum confinement effect results from the
vdW interfaces, which partially isolate electrons between
neighboring sheets. Another contributing factor may be from the
change of the dielectric environment, which is not included in the
$1/N^2$ law. It is apparent in Table \ref{table:1} and Fig.
\ref{fig:2} that the optical absorption peaks and exciton binding
energies are affected by the same mechanisms; they follow similar
scaling laws with the stacking layer number.

\begin{figure}
\includegraphics*[scale=1.00]{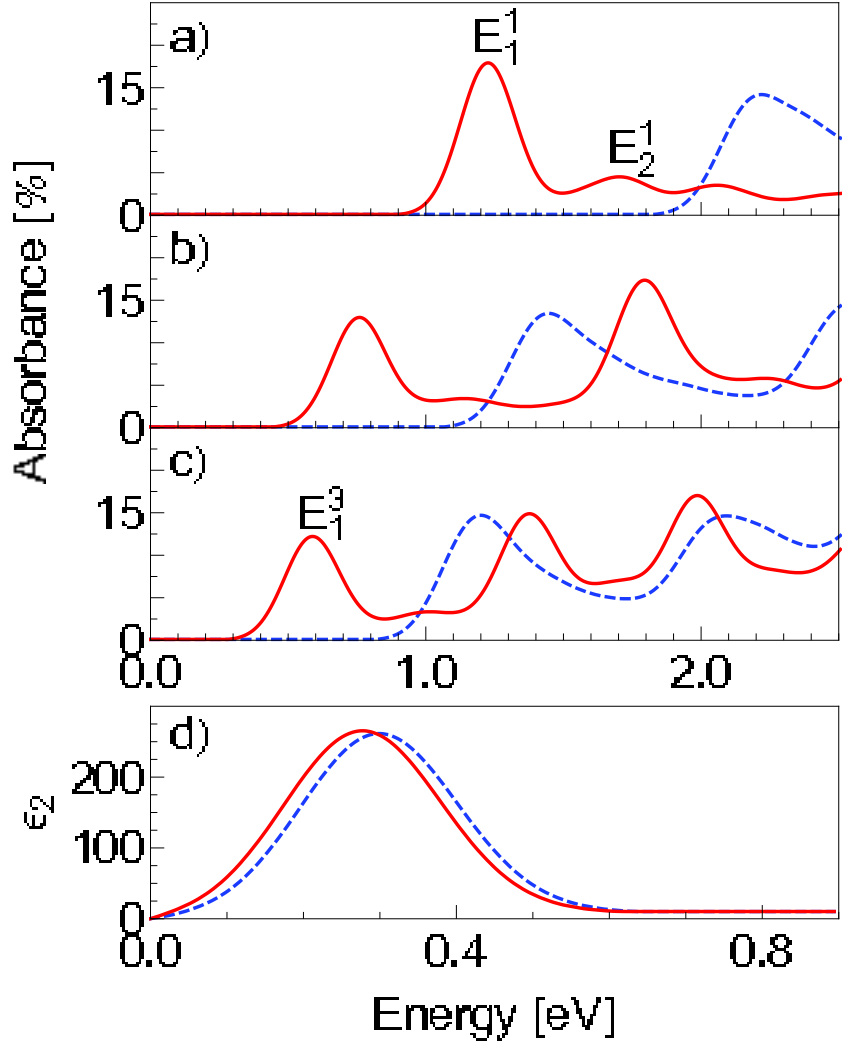}
\caption{(Color online) (a) to (d) Optical absorption spectra of
monolayer, bilayer, trilayer, and bulk phosphorene for the
incident light polarized along the x (armchair) direction. The
single-particle optical absorption are presented by dashed lines
while those spectra with \emph{e-h} interaction included are
presented by solid lines. We employ a 0.1 eV Gaussian smearing in
these plots. } \label{fig:3}
\end{figure}

The optical absorption spectra of monolayer, bilayer, trilayer and
bulk phosphorene structures are presented in Fig. \ref{fig:3} for
incident light polarized along the armchair ($\Gamma-X$)
direction. In all of the studied few-layer structures (Figs.
\ref{fig:3} (a) to (c)), excitonic effects substantially reshape
optical spectra; all of the main optical features are dominated by
excitonic states. For example, in the monolayer structure, the
first absorption peak is located at 1.2 eV, which is a strongly
bound excitonic state with an 800-meV \emph{e-h} binding energy.
These exciton binding energies in phosphorene are comparable to
those found in other monolayer semiconductors and 1D
nanostructures \cite{MoS2-2, 2004catalin, wang, ribbon}. The
reduced dimensionality and depressed screening are primary factors
for fostering such enhanced excitonic effects.

Recently, a photoluminescence measurement of monolayer phosphorene
was performed \cite{2014ye}. The measured spectral peak position
(1.45 eV) is marked in Fig. \ref{fig:2}; it resides slightly above
our single-shot GW-BSE result. Self-consistently updating the
Green's function spectrum and dielectric function using the
$G_1W_1$ methodology yields an exciton energy of 1.4 eV, as shown
in Fig. \ref{fig:2}. However, it should be noted that extrinsic
factors may influence experimental data \cite{2014ye}, which may
account for some of the disagreement with our calculations.
Therefore, additional experimental results must be assessed before
further conclusions can be made.

\begin{table}
\caption{\label{tb1} Fitted parameters for band gaps, the first
optical absorption peak (``optical gap") and exciton binding
energy of few-layer black phosphorus according to the formula
$A/N^{\alpha}+B$.}
\begin{center}
\begin{tabular}{ccccccccc}
\hline \hline

 & DFT/PBE & GW & Optical Gap & Binding Energy \\
\hline
 $\alpha$ & 0.85 & 0.73 & 0.96 & 0.53 \\
 A (eV) & 0.79 & 1.70 & 0.87 & 0.83 \\
 B (eV) & 0.04 & 0.30 & 0.28 & 0.03 \\
\hline \hline
\end{tabular}
\end{center}
\label{table:1}
\end{table}

In the bulk limit of black phosphorus, the optical absorption
spectrum is nearly unchanged by the inclusion of \emph{e-h}
interactions (Fig. \ref{fig:3} (d)). Our simulation estimates the
upper limit of the exciton binding energy as 30 meV, which is
similar to those in other bulk semiconductors \cite{2000Rohlfing}.
However, this is surprisingly different from similar layered
materials. For example, bulk hexagonal BN possesses a significant
exciton binding energy of 600 meV \cite{BN-1}. We attribute the
small excitonic effects in bulk phosphorous to its stronger
interlayer interactions, which is exhibited by its sizable
interlayer band dispersion \cite{struc-2}, making it a true
three-dimensional material unlike other, more weakly coupled layer
structures. We have also compared the electronic charge
distributions of monolayer and bulk phosphorene. The interlayer
interaction results in the reduction of interlayer distance and
substantial out-plane features (see the details in the
supplementary documents). This interlayer interaction and
corresponding coupling reduce the perpendicular quantum
confinement, resulting smaller band gaps and weaker excitonic
effects.

\begin{figure}
\includegraphics*[scale=0.60]{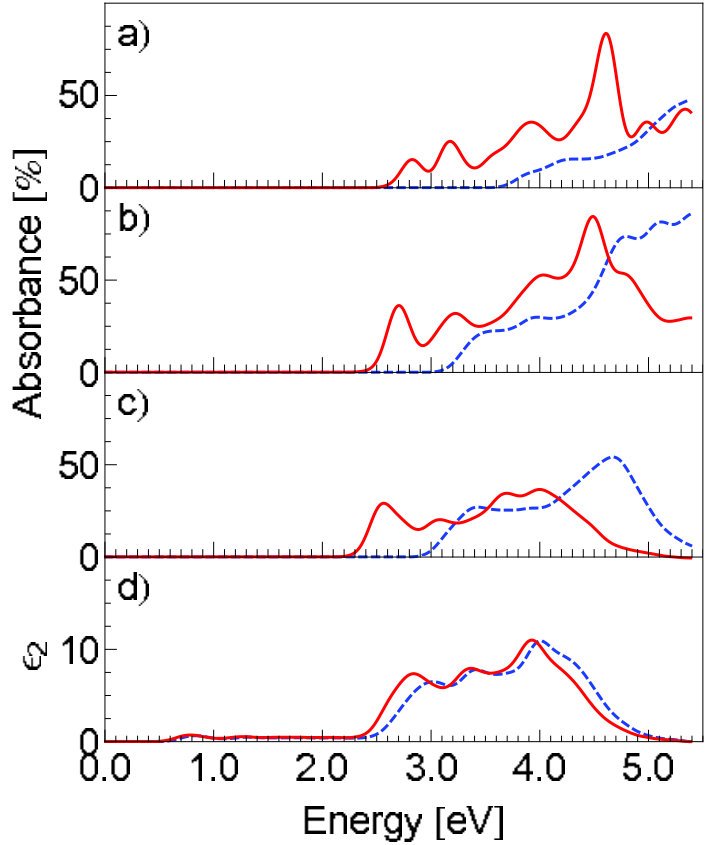}
\caption{(Color online) (a) to (d) Optical absorption spectra of
monolayer, bilayer, trilayer, and bulk phosphorene for the
incident light polarized along the y (zigzag) direction. The
single-particle optical absorption are presented by dashed lines
while those spectra with \emph{e-h} interaction included are
presented by solid lines.} \label{fig:4}
\end{figure}

The optical absorption spectra for light polarized along the
zigzag direction of few-layer black phosphorous is presented in
Fig. \ref{fig:4}. These host profoundly-distinct absorption
energy-ranges from the armchair spectra; the prominent absorption
features begin at higher energies - near 2.8 eV. Thus monolayer
phosphorene strongly absorbs armchair-polarized light with
energies between 1.1 eV and 2.8 eV and is transparent to
zigzag-polarized light in the same energy range; this phenomenon
is the result of selection rules associated with the symmetries of
this anisotropic material. Phosphorene is thus a natural optical
linear polarizer, which can be used in liquid-crystal displays,
three-dimensional visualization techniques, (bio)-dermatology, and
in optical quantum computers \cite{2009opt, 2001nat}. Furthermore,
the polarization energy window is tunable through a wide range.
Comparing Figs. \ref{fig:3} and \ref{fig:4}, the high-end of the
polarization window is nearly fixed at 2.8 eV, while the low-end
can be reduced from 1.1 eV down to 300 meV, by adjusting the layer
stacking number. This frequency range is very exploitable for
applications - it covers the infrared and near-infrared regimes.
Finally, this anisotropic optical response can be used to identify
the orientations of few-layer black phosphorous in experiments.

\begin{figure}
\includegraphics*[scale=0.40]{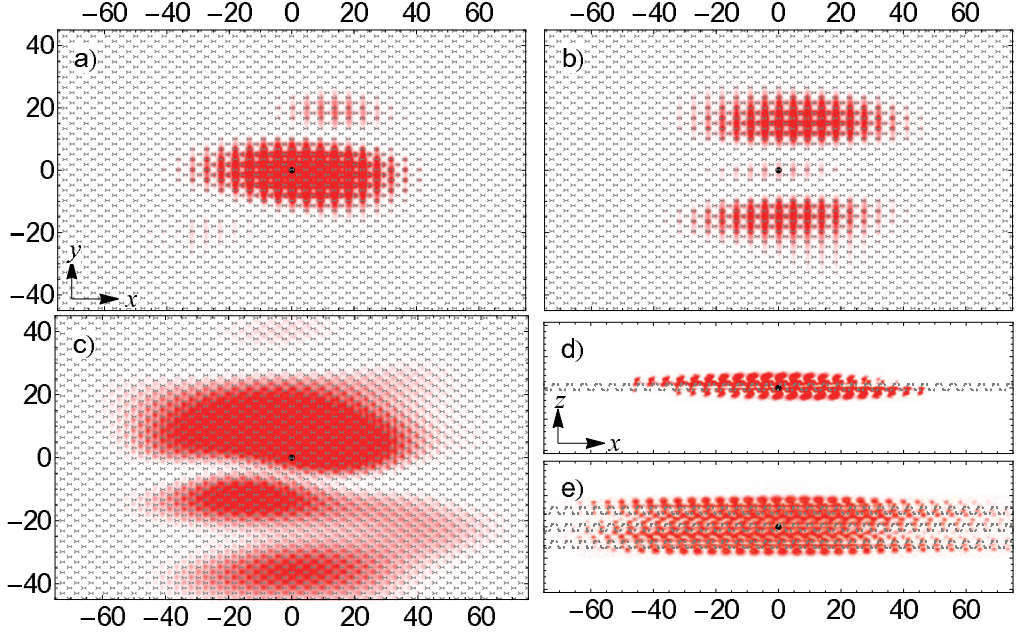}
\caption{(Color online) (a) and (b) Top views of the square of the
electron wavefunctions of the first and second bound excitons in
monolayer phosphorene (marked as $E^1_1$ and $E^1_2$ in Fig. 3
(a)). (c) Top view of the electron wavefunction of the first bound
exciton in bilayer phosphorene (marked as $E^3_1$ in Fig. 3 (c)).
(d) and (e) Side view of the wavefunctions in (a) and (c)
respectively. The hole, represented by a black dot, is fixed at
the origin. Lines representing the atomic bonds are superimposed.
The scale is in angstroms.} \label{fig:5}
\end{figure}

We have plotted wave functions of typical bright excitons in Fig.
\ref{fig:5}. An overall character of these excitons is that their
spatial distribution of wave functions is anisotropic and, in
particular, is extended along the armchair direction. This is
consistent with the anisotropic band dispersion shown in Fig.
\ref{fig:1} (b). These excitons form striped patterns, similar to
those found in bundles of nanowires. This is due to the fact that
the near-isotropic binding Coulomb interaction pulls on electrons
that are mobile only along one direction. Interestingly, the
optical activities of these plotted excitons are strongly
correlated with their spacial anisotropy; they are optically
bright only for the incident light polarized along the extended
direction (armchair direction) of their wave function, if we
compare Figs. \ref{fig:3}, \ref{fig:4}, and \ref{fig:5}.

If we compare wavefunctions of the first two bright excitons
($E^1_1$ and $E^1_2$ marked in Fig. \ref{fig:3} (a)) from the same
series in monolayer phosphorene (Figs. \ref{fig:5} (a) and (b)),
we see that their respective distributions resemble one another
both perpendicular to the structure and along the stripe direction
(armchair axis). It should be noted that the second excitonic
state ($E^1_2$) has a clear nodal structure perpendicular to the
stripe direction (along the zigzag axis), though one expects a 1D
exciton to exhibit nodes along the stripe direction. The wave
function of the first bright exciton ($E^3_1$) of trilayer
phosphorene is plotted in Figs. \ref{fig:5} (c) and (e). Although
the hole is fixed in one layer, electrons are distributed on both
layers. Thus the spatial extent of, and the interactions between
the excitons can also be controlled by the layer stacking number.

Finally, we assess the dependence of the many-electron effects on
the interlayer distance and dimensionality. Our studies show that
the interlayer distance slightly changes from bilayer to bulk
black phosphorus (less than 0.8 $\%$). Meanwhile, QP band gaps and
intralayer excitons of suspended few-layer phosphorene (which is
treated with a Coulomb truncation) are not very sensitive to the
change of interlayer distances. For example, the band gap of
bilayer phosphorene varies by less than 60 meV for the change of
interlayer distance of $\pm 0.5 \AA$. These many-electron effects
are rooted in the vast vacuum that surrounds the isolated system,
and are consequently not significantly affected by small changes
of the interlayer distance. The layers of bulk black phosphorous,
on the other hand, do not interface with a vacuum, and their
excited state properties are sensitive to the interlayer distance.
A small change of ($\pm 0.5 \AA$) in the interlayer distance can
shift the band gap and exciton energy by 150 meV, which is
significant by comparison to the 300-meV band gap. (Additional
details can be found in the supplementary document.) Obviously, if
the thickness of few-layer black phosphorus is larger than the
characteristic size of QPs or excitons, many-electron effects
shall smoothly evolve to the bulk limit. Based on sizes of
excitons plotted in Fig. \ref{fig:5}, we estimate the critical
thickness is around 10 nm, which is roughly around 20 layers.

In conclusion, first-principles simulations have been performed to
determine the defining properties of few-layer black phosphorus
and their exciting potential applications. Enhanced many-electron
effects are essential in shaping their band gaps and optical
responses because of anisotropic band dispersions and the
effectively quasi-1D nature. In particular, our discovered
uniquely anisotropic optical response with \emph{e-h} interactions
included makes phosphorene ideal for linear optical polarizers,
covering the infrared and a part of visible light regime of broad
interest. Finally, we show that all these properties can be
efficiently tuned by the stacking layer number, which is
potentially useful for device design.

We are supported by NSF Grant No. DMR-1207141. We acknowledge Erik
Henriksen and Ruixiang Fei for fruitful discussions. The
computational resources have been provided by Lonestar of Teragrid
at the Texas Advanced Computing Center. The ground state
calculation is performed by the Quantum Espresso
\cite{2009Giannozzi}. The GW-BSE calculation is done with the
BerkeleyGW package \cite{2012Deslippe}.

\end{document}